\documentclass[fleqn,10pt]{wlscirep}
\usepackage[utf8]{inputenc}
\usepackage[T1]{fontenc}
\usepackage{listings}
\definecolor{codegreen}{rgb}{0,0.6,0}
\definecolor{codegray}{rgb}{0.5,0.5,0.5}
\definecolor{codepurple}{rgb}{0.58,0,0.82}
\definecolor{backcolour}{rgb}{0.95,0.95,0.92}

\lstdefinestyle{mystyle}{
    backgroundcolor=\color{backcolour},   
    commentstyle=\color{codegreen},
    keywordstyle=\color{magenta},
    numberstyle=\tiny\color{codegray},
    stringstyle=\color{codepurple},
    basicstyle=\ttfamily\footnotesize,
    breakatwhitespace=false,         
    breaklines=true,                 
    captionpos=b,                    
    keepspaces=true,                 
    numbers=left,                    
    numbersep=5pt,                  
    showspaces=false,                
    showstringspaces=false,
    showtabs=false,                  
    tabsize=2
}
\usepackage{subcaption}

\title{QeMFi: A Multifidelity Dataset of Quantum Chemical Properties of Diverse Molecules}

\author[1,*]{Vivin Vinod}
\author[1]{Peter Zaspel}
\affil[1]{School of Mathematics and Natural Science, University of Wuppertal, 42119 Wuppertal, Germany}

\affil[*]{corresponding author(s): Vivin Vinod (vinod@uni-wuppertal.de)}

\begin{abstract}
Progress in both Machine Learning (ML) and Quantum Chemistry (QC) methods have resulted in high accuracy ML models for QC properties. Datasets such as MD17 and WS22 have been used to benchmark these models at some level of QC method, or fidelity, which refers to the accuracy of the chosen QC method. 
Multifidelity ML (MFML) methods, where models are trained on data from more than one fidelity, have shown to be effective over single fidelity methods. Much research is progressing in this direction for diverse applications ranging from energy band gaps to excitation energies. One hurdle for effective research here is the lack of a diverse multifidelity dataset for benchmarking. 
We provide the Quantum chemistry MultiFidelity (QeMFi) dataset consisting of five fidelities calculated with the TD-DFT formalism. The fidelities differ in their basis set choice: STO-3G, 3-21G, 6-31G, def2-SVP, and def2-TZVP. QeMFi offers to the community a variety of QC properties such as vertical excitation properties and molecular dipole moments, further including QC computation times allowing for a time benefit benchmark of multifidelity models for ML-QC.
\end{abstract}
\begin{document}

\flushbottom
\maketitle

\thispagestyle{empty}

\section*{Background \& Summary}
Recent developments in the field of machine learning (ML) for quantum chemistry (QC) have significantly changed the landscape of research and discovery in QC properties \cite{Westermayr2020review, Sergei21_Chem_review_NNML, dral2020quantum, dral21a} with significant reduction in the time to predict QC properties once an ML model has been trained. For such models, the protocol often involves testing them against some benchmark datasets such as the MD17 \cite{chmiela2017}, QM7 \cite{blum_QM7_dataset}, or the QM9 dataset \cite{Rudd2012_QM9_dataset, ramakrishnan2014_QM9_dataset}. Recently, the WS22 database was released with a collection of Wigner Sampled geometries of 10 diverse molecules \cite{ws22_data_paper, ws22_dataset_zenodo}. With varied chemical complexity and number of atoms, the WS22 datasets provides a collection of QC properties for these molecules calculated at one level of theory, or \textit{fidelity}. It was also shown that for this collection of molecules the use of ML methods is indeed challenging due to the wider chemical space that the geometries cover \cite{ws22_data_paper, hou_2023_KREGpKREG_ws22}. 

Multifidelity methods harnessing inherent QC hierarchies to cancel out errors across different numerical QC methods have since superseded the single fidelity ML methods. These methods include $\Delta$-ML \cite{Ramakrishnan2015} based models such as hierarchical machine learning \cite{dral2020hierarchical}, multifidelity machine learning (MFML) \cite{zasp19a, vinod23_MFML}, and optimized MFML (o-MFML) \cite{vinod_2024_oMFML}. Certain other flavors of ML using multifidelity data have been proposed and tested, including multi-task Gaussian processes treating the different fidelities as interdependent tasks \cite{fisher2024multitask, ravi2024multifidelity}. Multifidelity methods have been applied to predicting diverse QC properties such as band gaps in solids, excitation energies, and atomization energies of various molecules \cite{Pilania2017, patra2020multi, zasp19a, vinod23_MFML}.

Several QC datasets have been generated for the general work of ML for QC, some consisting of multiple fidelities of data. Some of these include ground state energies and electronic spectra data computed at DFT level of theory and some semi-empirical levels of theory hosted in the bigQM7$\omega$ \cite{Smith_Zubatyuk_Nebgen_Lubbers_Barros_Roitberg_Isayev_Tretiak_2020_dataset} dataset.
The PubCheMQC project presented a large database of electronic structure properties calculated with the DFT formalism using two distinct basis sets for use in training ML models \cite{Nakata_Shimazaki_2017_dataset}.
The QM8 \cite{Ramakrishnan_Hartmann_Tapavicza_von_Lilienfeld_2015_dataset} dataset records electronic spectra properties with DFT formalism for over 20k geometries of small organic molecules sampled from the larger QM9 dataset \cite{Rudd2012_QM9_dataset, ramakrishnan2014_QM9_dataset}. Ref.~\cite{patra2020multi} introduced a multifidelity dataset of 358 polymer bandgaps for benchmarking use with multifidelity co-kriging methods. The ANI-1x and ANI-1ccx datasets provide a rich multifidelity dataset of around 5 million data points with HF, MP2, and NNPO-CCSD(T) energies and forces primarily created for the training of the ANI-1x potential \cite{Smith_Zubatyuk_Nebgen_Lubbers_Barros_Roitberg_Isayev_Tretiak_2020_dataset, Smith_Nebgen_Lubbers_Isayev_Roitberg_2018_anipotential}. VIB5 is yet another multifidelity dataset with ab initio quantum chemical properties including potential energy surfaces for five molecules with MP2, HF, and CCSD(T) levels of theory for different basis set sizes \cite{Zhang_Zhang_Owens_Yurchenko_Dral_2022_dataset}. The QM7b dataset consists of multifidelity atomization energies computed at the MP2, HF, and CCSD(T) fidelities for three distinct basis set sizes with 7,211 small-to-medium sized molecules \cite{blum_QM7_dataset, montavon2013machine}. Ref.~\cite{Fediai_Reiser_Peña_Friederich_Wenzel_2023_dataset} offers orbital energies of 134k molecules for PBE and GW fidelities. The MultiXC-QM9 is an elaborate dataset of diverse QC properties such as reaction energies which are computed with the DFT formalism using 76 different functionals for three basis sets \cite{Nandi_Vegge_Bhowmik_2023_dataset}. 
All of these above mentioned datasets have important use-cases for multifidelity machine learning methods and related benchmarks. However, none of these datasets offer the QC compute time for the different fidelities present. That is, although multifidelity models can be created and benchmarked in terms of model error, it is not possible to use these datasets to perform time-cost benchmarks for multifidelity models. Since the entire conceptualization of multifidelity methods such as $\Delta$-ML \cite{Ramakrishnan2015} or MFML \cite{zasp19a} is to reduce the cost of generating training data, this key factor is necessary to meaningfully benchmark these methods. A mere model accuracy benchmark is insufficient.

To unify the research in this rapidly developing field of multifidelity methods, it becomes necessary to present to the community a diverse collection of multifidelity data over a range of molecular complexity which also includes the time-cost of the QC calculations. Building up on existing datasets is preferred in such a scenario to prevent redundant calculations and geometry generation. After all, the entire point of a multifidelity method is to reduce compute cost and resource usage in discovery and research. In interest of such an approach, the WS22 database \cite{ws22_dataset_zenodo} was chosen to be the collection of geometries. In addition to being a collection of molecules that are chemically complex with distinct conformers, the molecule in this dataset also cover a wide range of the quantum chemical configuration space in contrast to other datasets such as MD17. The presence of flexible functional groups make the geometries, and by extension, the QC properties, of this dataset challenging for ML models to learn \cite{ws22_data_paper}. These features make this collection the preferred choice to generate multifidelity data. For each of the molecules of increasing size and chemical complexity, this dataset offers 120,000 geometries. This creates a vast dataset collection of diverse geometries covering various conformers of the different molecules. In total there are around 1 million geometries in the WS22 database. Performing multifidelity QC calculations for such a vast number of geometries is not feasible. It is more realistic and computationally feasible to produce a multifidelity dataset for a portion of the geometries of the WS22 database. Therefore, for each of the molecules in the WS22 database, 15,000 geometries were evenly sampled, for a total of $9\times 15,000=135,000$ geometries, and the multifidelity QC calculations performed for these. 

This dataset is provided to the ML-QC community under the name QeMFi (\textbf{Q}uantum Ch\textbf{e}mistry \textbf{M}ulti\textbf{Fi}delity) dataset \cite{vinod_2024_QeMFi_zenodo_dataset}. A detailed description of the geometry sampling, data generation procedure, the fidelities, and the technical details of the QeMFi dataset are provided in the following section. In addition, scripts to generate two multifidelity models from ref.~\cite{vinod_2024_oMFML}, namely, MFML and optimized MFML (o-MFML) are provided. These ML methods are discussed in the Methods section of this work.
Scripts to assess time benefit of multifidelity methods are also included. This makes it easy for future research in the multifidelity methods to establish a clear time benefit for these models over standard single fidelity ML methods.
The diverse collection of molecules in QeMFi along with their multifidelity properties, provides a challenging dataset for the domain of ML in QC. Due to the large number of multifidelity data points along with their QC time-costs and easily usable associated scripts, we believe that QeMFi is a significant collection that will help push the boundaries of multifidelity methods for ML in QC properties enabling meaningful time-cost assessments for these methods. 

\section*{Methods}
The original WS22 database includes the following molecules (in increasing order of number of atoms):
\begin{enumerate}
    \item urea
    \item acrolein
    \item alanine
    \item 2-(methyliminomethyl)phenol (SMA)
    \item 2-nitrophenol
    \item urocanic acid
    \item 4-(dimethylamino)benzonitrile (DMABN)
    \item thymine
    \item 4-(2-hydroxybenzylidene)-1,2-dimethyl-1H-imidazol-5(4H)-one (o-HBDI)
\end{enumerate}
In addition to these molecules, toluene is also included to compare with the MD17 \cite{chmiela2017} database. Since toluene consists of a single conformer and was only introduced in WS22 for comparison to existing datasets such as MD17, this molecule was not included while generating the QeMFi dataset.
The original WS22 database was first generated as reported extensively in ref.~\cite{ws22_data_paper}. The pipeline involves optimized equilibrium geometries identification for the different conformations of the molecule with DFT \cite{Quin_MP2_DFT_theory_2005, TD_DFT_Runge_1984}. Following this, the respective Wigner Sampling is carried out from ground state (S0) and/or excited state (S1) minima. 
For these, the geometries are subsequently interpolated by finding on a Riemann manifold, an optimized geodesic curve. The metric for this is defined by a redundant internal coordinate functions \cite{zhu_geodesicinterpol_2019}. 
In the original WS22 database provided in ref.~\cite{ws22_dataset_zenodo}, this results in a little over 1 million samples across 9 molecules with various properties calculated at the TD-DFT level of theory using the PBE0/6-311 G* functional and basis set combination \cite{ws22_data_paper}. 

\subsection*{Data Sampling and Quantum Chemistry Calculations}
To build the QeMFi dataset from the WS22 database, 15,000 geometries were sampled from the original 120,000 geometries for each of these molecules. 
For each of the nine molecules from the WS22 database, 15,000 geometries were evenly sampled from the original 120,000 geometries. To achieve this, every $8^{\rm th}$ geometry for each molecule was selected from the WS22 database resulting in a total of $9\times 15,000=135,000$ point geometries for QeMFi. An even sampling of the original dataset ensures that there are sufficient geometries from all conformations of the molecule. Once these geometries were sampled, they were used to perform point calculations for the QC properties.

\begin{figure}[htb!]
    \centering
    \includegraphics[width=\linewidth,trim = 0cm 3.5cm 0cm 0cm, clip]{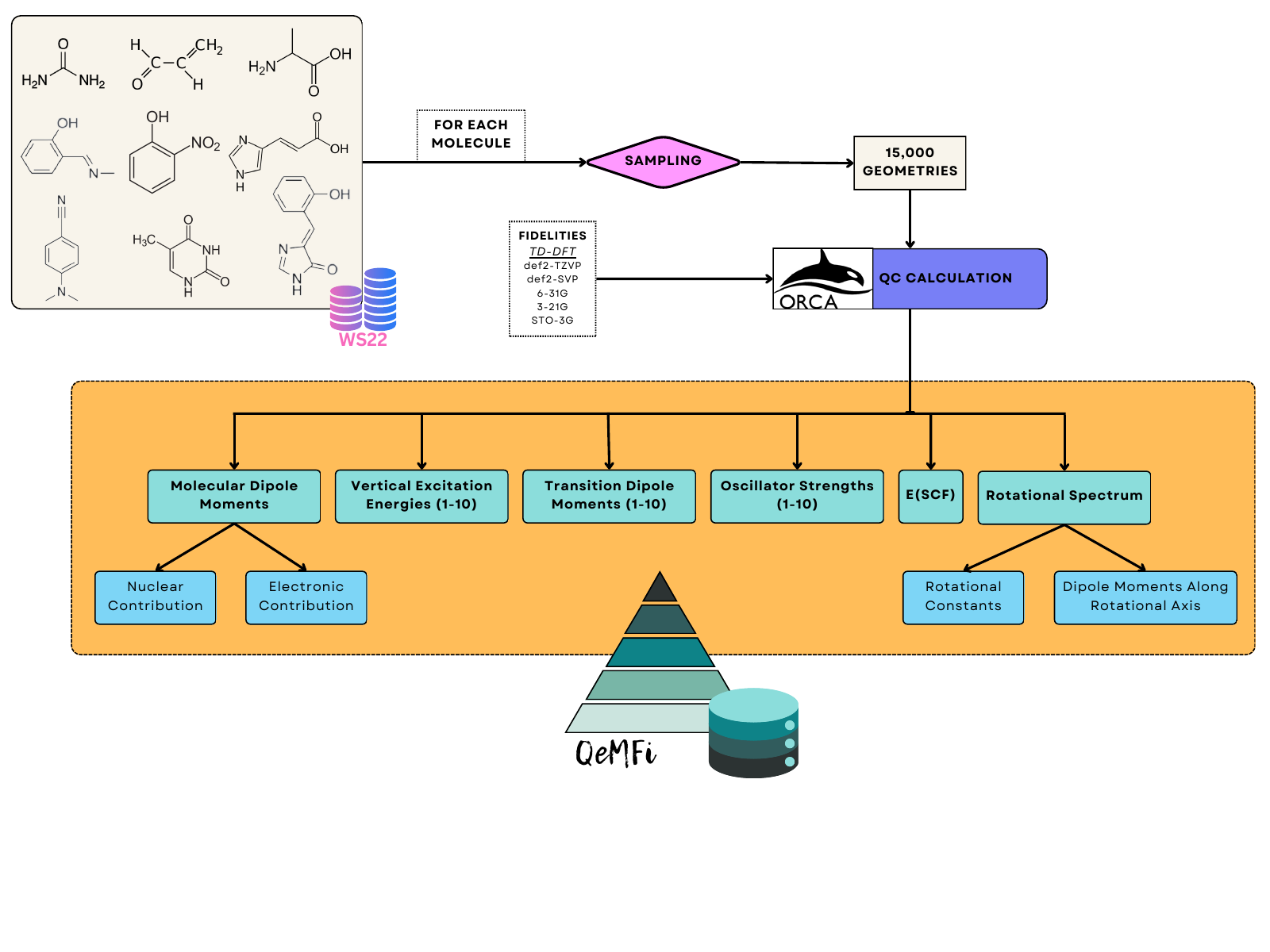}
    \caption{The workflow of generating the QeMFi dataset by sampling from the WS22 database. 15,000 geometries are used for each molecule resulting in a total of 135,000 single point geometries. For each of these, multiple QC properties are calculated at DFT level of theory with varying basis set sizes to create the diverse multifidelity dataset.}
    \label{fig_workflow}
\end{figure}

All QC calculations were performed with the ORCA(5.0.1) QC package \cite{nees20a}. From these calculations, a diverse set of QC properties were extracted including information of the vertical excitation states such as energies and oscillator strengths. 
QC calculations were performed at the TD-DFT level of theory with the CAM-B3LYP functional. 
For each geometry, five fidelities were calculated. These fidelities are the basis set choice of increasing size. In increasing hierarchy of the fidelity, these are: STO-3G, 3-21G, 6-31G, def2-SVP, and def2-TZVP.  In the rest of the document, for the most part, these are referred to by their short-hand, i.e., STO3G up to TZVP.
The \texttt{TightSCF} keyword was employed to ensure energy convergence of the order of $10^{-9}$ a.u.~for each calculation. Resolution of Identity approximation (RIJCOSX) was employed in order to speed up the excitation energy calculations. For any calculation, the maximum memory usage was limited to 2.0 GB. 
In practice, the ORCA calculations did not use this amount of memory. 
A total of 10 vertical excitation energies were calculated with each fidelity for each geometry. 
The complete workflow of the dataset generation process is pictorially depicted in Figure \ref{fig_workflow}.

A note on the calculations being restricted to DFT methods is to be made here. Since QeMFi is a benchmark dataset and not a high accuracy model training dataset, the cost of generating a costlier dataset, say at coupled cluster level of theory, was considered to be excessive. The aim of this dataset is to present a diverse collection of QC properties based on a set of complex molecules which can be used to uniformly assess MFML methods. Therefore, the QC properties are calculated only with DFT methods and higher accuracy methods such as the gold standard CCSD(T) are not considered.

The list of available multifidelity properties is given in Table \ref{tab_properties_list}. 
The Cartesian coordinates and atomic numbers are taken from the WS22 database. The SCF ground state energies are reported in Hartree units. 
The first 10 vertical excitation energies are provided in $\rm cm^{-1}$ with their corresponding oscillator strengths and transition dipole moments (in a.u.). 
The molecular dipole moments are also a property included in the QeMFi dataset with both the nuclear and electronic contributions being separately cataloged in atomic units (a.u.). 
The rotational spectrum data is also included in the form of Rotational Constants (in $\rm cm^{-1}$) and the total molecular dipole moments (in a.u.) aligned along rotational axes.

As an important contribution to the multifidelity research, the average time to run the QC computations are also provided for each molecule for each fidelity. This information can be further used to benchmark multifidelity models across QC properties as was shown in ref.~\cite{vinod23_MFML} for excitation energies of arenes. The notion here is to assess the error of a model with respect to the time it takes to generate a training set for that specific model. Such an analysis of the model, as shown in ref.~\cite{vinod23_MFML}, shows the true time benefit of multifidelity models, in this specific case, for MFML. In the QeMFi dataset, the time to run an ORCA calculation for a specific fidelity is provided for each molecule in units of seconds.
The parallelization of numerical codes is a challenge unto itself and usually comes with artifacts that arise due to the form of parallelization. This results in non-uniform calculations schemes across basis set sizes and molecules. In order to enforce consistency in the calculation times, ORCA calculations were run on a single compute core for 10-evenly sampled geometries of each molecule. The calculation times as returned by the ORCA software are then averaged over these 10 geometries and reported for each fidelity for each molecule. Thus, the time for a single-core calculation of each fidelity for each molecule is provided in units of seconds to benchmark the time-benefit of multifidelity models against single fidelity models.
This diverse collection of QC properties is made available for $9\times 15,000=135,000$ geometries across five different fidelities providing ample room for development and benchmarking of MFML methods and models. 

\subsection*{Machine Learning Methods}
Since this data publication is made along with scripts that enable benchmarking of multifidelity methods for ML-QC, a short digression is made here to discuss the ML methods used in these scripts. The choice of ML method is kernel ridge regression (KRR). The use of diverse ML methods for QC is out of scope of this data publication.
The accompanying benchmarking scripts use the MFML and o-MFML methods as described in ref.~\cite{zasp19a,vinod23_MFML,vinod_2024_oMFML}. These two methods are also used to provide technical benchmarks for the QeMFi dataset including time-benefit of the multifidelity methods over single fidelity methods. Thus, MFML and o-MFML also briefly explained here for the sake of completeness.

\subsubsection*{Kernel Ridge Regression}
Let $\mathcal{T}:=\{(\boldsymbol{X}_i,E_i)\}_{i=1}^{{N}_{\rm train}}$ be a training set of size ${N}_{\rm train}$ for molecular descriptors $\boldsymbol{X}_i$ and their corresponding QC property, $E_i$, be given. 
A KRR model for the prediction of the QC property $E_i$, for an unseen query descriptor $\boldsymbol{X}_q$, is denoted by
\begin{equation}
    P_{\rm KRR}\left(\boldsymbol{X}_q\right) := \sum_{i=1}^{N_{\rm train}} \alpha_i k\left(\boldsymbol{X}_q,\boldsymbol{X}_i\right)~,
    \label{eq_KRR_def}
\end{equation}
where $k$ is the kernel function. The various coefficients, $\boldsymbol{\alpha}$, are trained by solving $(\boldsymbol{K}+\lambda \boldsymbol{I}) \boldsymbol{\alpha} = \boldsymbol{E}$, where $\boldsymbol{K} = \left(k(\boldsymbol{X}_i,\boldsymbol{X}_j)\right)_{i,j=1}^{{N}_{\rm train}}$ is the kernel matrix, $\boldsymbol{I}$ the identity matrix, $\boldsymbol{E} = \left(E_1, E_2, \ldots, E_{{N}_{\rm train}}\right)^T$ is the vector of QC properties from the training set and $\lambda$ a regularization parameter. 
This work uses two kernels while executing KRR. The first is the Mat\'ern Kernel of first order with the discrete L-2 norm:
\begin{equation}
    k\left(\boldsymbol{X}_i,\boldsymbol{X}_j\right) = \exp{\left(-\frac{\sqrt{3}}{\sigma}\left\lVert \boldsymbol{X}_i-\boldsymbol{X}_j\right\rVert_2^2\right)}\cdot\left(1+\frac{\sqrt{3}}{\sigma}\left\lVert \boldsymbol{X}_i-\boldsymbol{X}_j\right\rVert_2^2\right)~,
    \label{eq_matern}
\end{equation}
where $\sigma$ denotes a length scale hyperparameter that determines the width of the kernel.
The second kernel used is the Laplacian kernel which is given as
\begin{equation}
    k\left(\boldsymbol{X}_i,\boldsymbol{X}_j\right) = \exp{\left(-\frac{\left\lVert \boldsymbol{X}_i-\boldsymbol{X}_j\right\rVert_1}{\sigma}\right)}~.
    \label{eq_Laplacian}
\end{equation}

\subsubsection*{Multifidelity Machine Learning}
Let the training set for data at some fidelity $f$ be $\mathcal{T}^{(f)}:=\left\{\left(\boldsymbol{X}^{(f)}_i,y_i^{(f)}\right) \right\}_{i=1}^{N^{(f)}}$. Let an ordered hierarchy of fidelities be given as $f\in\{1,2,\ldots F-1,F\}$ where $F$ is the most accurate fidelity and is called the target fidelity.
In ref.~\cite{zasp19a, vinod_2024_oMFML} it was shown that a MFML model can be built:
\begin{equation}
    P_{\rm MFML}^{(F,\eta_F;f_b)}\left(\boldsymbol{X}_q\right) := \sum_{\boldsymbol{s}\in \mathcal{S}^{(F,\eta_F;f_b)}} \beta_{\boldsymbol{s}} P^{(\boldsymbol{s})}_{\rm KRR}\left(\boldsymbol{X}_q\right)~,
    \label{eq_MFML_linearsum}
\end{equation}
where, $f_b<F$ is called the baseline fidelity, that is the cheapest fidelity used in the multifidelity scheme. The linear combination is performed for the various sub-models built with different fidelities. Each sub-model is denoted by $\boldsymbol{s}=(f,\eta_f)$, where $2^{\eta_f}=N_{\rm train}^{(f)}$. That is, a sub-model is identified by the fidelity and the number of training samples used at that fidelity. 
The sub-models are chosen based on a scheme discussed extensively in ref.~\cite{vinod_2024_oMFML} and form the set of sub-models given by $\mathcal{S}^{(F,\eta_F;f_b)}$. 
The different $\beta_{\boldsymbol{s}}$ are the coefficients of the linear combination. 
Based on previous research \cite{zasp19a}, the $\beta_{\boldsymbol{s}}$, are set in for MFML as follows:
\begin{equation}
    \beta_{\boldsymbol{s}}^{\rm MFML} = \begin{cases}
        +1, & \text{if } f+\eta_f = F+\eta_F\\
        -1, & \text{otherwise}
    \end{cases}~.
    \label{eq_MFML_beta_i}
\end{equation}

\subsubsection*{Optimized MFML}
The optimized MFML (o-MFML) method was introduced in ref.~\cite{vinod_2024_oMFML} where the combination of the different sub-models built on different fidelities is optimized on a holdout validation set of reference properties. Such an o-MFML was shown to improve the accuracy of prediction for atomization energies and excitation energies across diverse molecules.
o-MFML rewrites Eq.~\eqref{eq_MFML_linearsum} by considering values of the coefficients different from that given in Eq.~\eqref{eq_MFML_beta_i}. This is done by optimizing the coefficients on a holdout validation set. In this sense, the different $\beta_{\boldsymbol{s}}$ are treated as hyperparameters and this entire process is akin to a hyper-parameter optimization routine which is commonly used in ML methods.

One can define the validation set for such optimization as
$\mathcal{V}^F_{\rm val}:=\{(\boldsymbol{X}_q^{\rm val},E^{\rm val}_q)\}_{q=1}^{N_{\rm val}}$. 
This validation-test set approach is common in ML techniques. The hyperparameters are optimized on the validation set and the model is evaluated on the test set.

For a target fidelity $F$, with $N^{(F)}=2^{\eta_F}$ training samples at the target fidelity, for a baseline fidelity $f_b$, the o-MFML model is defined as 
\begin{equation}
    P_{\rm o-MFML}^{\left(F,\eta_F;f_b\right)}\left(\boldsymbol{X}_q\right) := 
    \sum_{\boldsymbol{s}\in \mathcal{S}^{(F,\eta_F;f_b)}}\beta_{\boldsymbol{s}}^{\rm opt}\cdot P^{(\boldsymbol{s})}_{\rm KRR} \left(\boldsymbol{X}_q\right)~,
    \label{eq_POM_def}
\end{equation}
where $\beta_{\boldsymbol{s}}^{\rm opt}$ are optimized coefficients.
This form of optimization is carried out by solving the following optimization problem:
$$
    \beta_{\boldsymbol{s}}^{\rm opt} = \arg\min_{\beta_{\boldsymbol{s}}} 
    \left\lVert \sum_{v=1}^{N_{\rm val}} \left(y_v^{\rm ref} - \sum_{\boldsymbol{s}\in S'} \beta_{\boldsymbol{s}}\cdot P^{(\boldsymbol{s})}_{\rm KRR}\left(\boldsymbol{X}_v\right)\right) \right\rVert_p 
$$
where one minimizes some $p$-norm on $\mathcal{V}^F_{\rm val}$. 
Based on ref.~\cite{vinod_2024_oMFML}, the ordinary least squares (OLS) method is used to optimize the coefficients which implicitly uses a 2-norm. 

\subsubsection*{ML Error Metric}
To technically verify the ML benchmarks performed in this work, the model error is computed with the mean absolute error (MAE). For a test set computed at the target fidelity, $\mathcal{T}^{F}_{test}:=\{(\boldsymbol{X}_i,y_i^{ref}\}_{i=1}^{N_{test}}$, this error is computed as 
\begin{equation}
    MAE:= \frac{1}{N_{test}}\sum_{i=1}^{N_{test}} \left\lvert y_i^{ref}-y_i^{ML} \right\rvert,
\end{equation}
where, $y_i^{ML}$ refers to the ML prediction of the $i$th geometry in the test set. The ML model could be either the single fidelity KRR or the MFML and o-MFML models described above. Learning curves, a depiction of MAEs as a function of the increasing number of training samples used at the target fidelity, are used as a visual metric for these models. The learning curves are shown for a 10-run average run with a shuffling of only the training data with the algorithm detailed in ref.~\cite{vinod_2024_oMFML}. It is to be noted that the MAE is only reported with respect to the target fidelity, as is common in multifidelity method assessments \cite{zasp19a, vinod23_MFML, vinod_2024_oMFML, fisher2024multitask}.

\section*{Data Records}\label{datarecords}
The various QC properties of the QeMFi dataset are stored in separate NumPy (v 1.26.4) \texttt{npz} files for each molecule. These \texttt{npz} files have a dictionary-like format allowing for each property to be accessed via its corresponding key denoted in Table \ref{tab_properties_list}. Each property itself is stored as a NumPy \texttt{ndarray} with the first dimension being 15,000 corresponding to the number of geometries. Thus, the QC properties can be accessed by querying the right ID. For example, the SCF ground state energies can be accessed with the key \texttt{`SCF'} returning a NumPy \texttt{ndarray} of size $15,000\times5$ where the second dimension of the array corresponds to the five fidelities used.
Similarly, one can access the QC computation times using the key \texttt{`t'} which results in a NumPy array of shape $(5,)$ corresponding to the five fidelities used. The compute times are stored in units of seconds.
An example script to accessing the QC properties is shown in Listing \ref{lst_codeaccess}.

The dataset itself is hosted on Zenodo at \href{https://doi.org/10.5281/zenodo.13925688}{https://doi.org/10.5281/zenodo.13925688} with a detailed README file documenting the key aspects of the data \cite{vinod_2024_QeMFi_zenodo_dataset}. The README also provides information on how to access the different properties using Python.
For the QeMFi dataset, the various scripts involved in generating the data, including ORCA input files and shell scripts to extract properties from the ORCA log files, are stored in the code repository that can be accessed at \href{https://github.com/SM4DA/QeMFi}{https://github.com/SM4DA/QeMFi}. 
In addition to these scripts, the code repository also contains Python scripts to perform multifidelity benchmarks on this dataset. These can be launched using the CLI and are a handy tool in setting benchmarks for this dataset using current state of art multifidelity methods. 

\section*{Technical Validation}\label{benchmark}
In order to verify that this form of sampling did in fact evenly cover the conformation space of each molecule, Uniform Manifold Approximation and Projections (UMAPs) \cite{umap} are studied herein. 
To further validate the QeMFi dataset and its use in benchmarking multifidelity methods, the MFML and o-MFML models prescribed in ref.~\cite{vinod_2024_oMFML} were tested in predicting ground state energies and the first vertical excitation energies. The multifidelity models are built for different baseline fidelities, which refers to the cheapest fidelity included in the model. For example, a baseline fidelity $f_b=$631G implies that the multifidelity model is built up of the fidelities 631G, SVP, and TZVP \cite{vinod_2024_oMFML}.
In addition to benchmarking the multifidelity models on properties of individual molecules, the models are also tested on using data from all the molecules of the dataset. For this purpose, the ground state energy of all molecules are used to train one single MFML and o-MFML model. This is then tested on predicting the ground state energies of all molecules. 

While these broad tests serve as a benchmark for multifidelity models on this dataset, the benchmarks of the other properties and molecules are not reported here. However, it is to be pointed out that the scripts provided can be readily used to generate benchmarks for these cases using standard ML methods such as learning curves. All learning curves are reported for a 10-run average, that is, for 10 random shuffling of the training set as directed in ref.~\cite{vinod_2024_oMFML}.

\subsection*{Conformation Space Coverage}
\begin{figure}[htb!]
    \centering
    \includegraphics[width=0.75\linewidth]{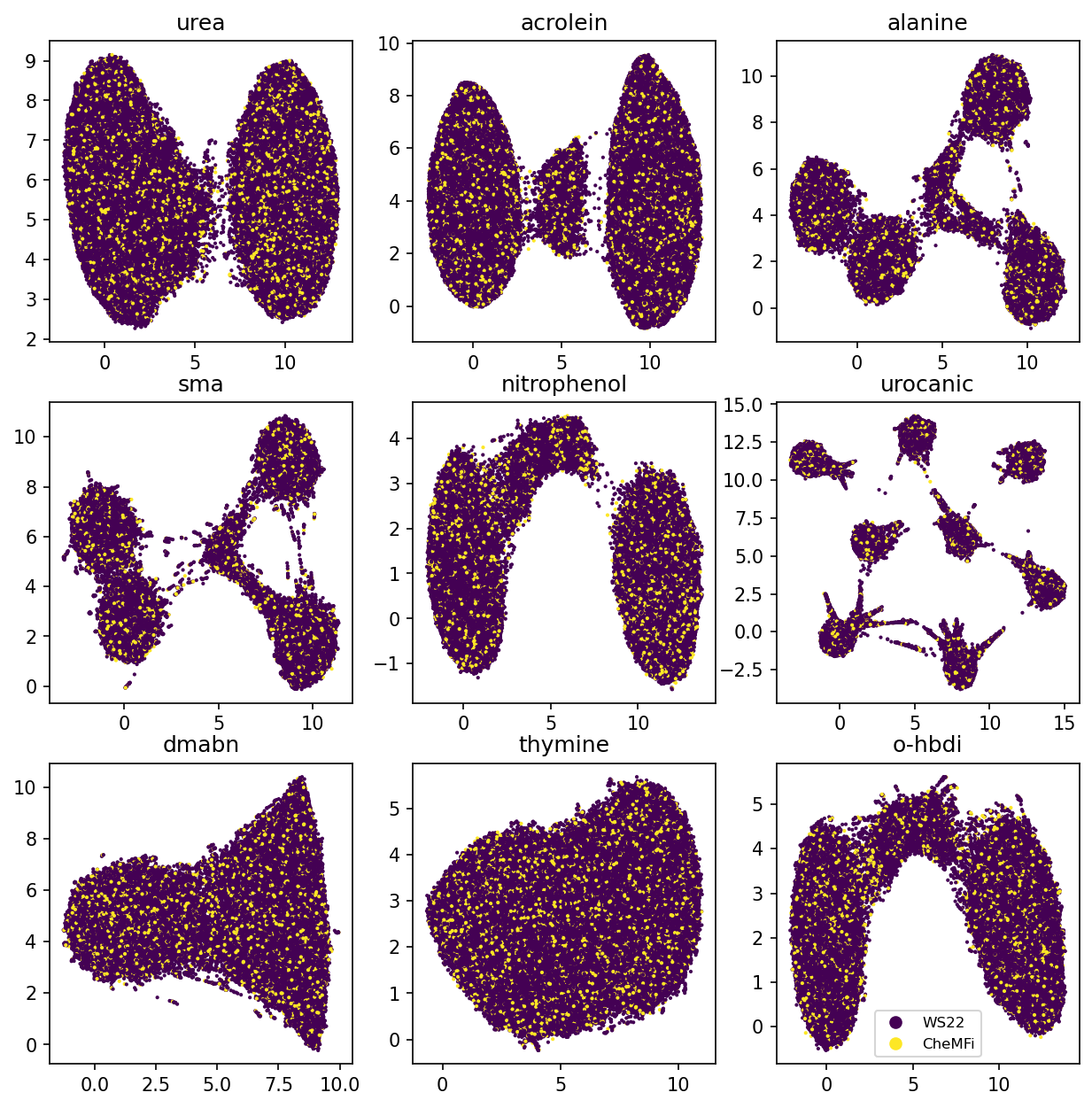}
    \caption{Scatter plots of UMAPs for the various molecules which compose the WS22 database. The UMAPs were generated for the unsorted Coulomb Matrix (CM) molecular descriptor for each molecule. The legend key indicates the geometries which are part of the WS22 and the QeMFi dataset respectively. For all molecules, it can be observed that the QeMFi dataset traverses the entirety of the configuration space that WS22 also covers.}
    \label{fig_UMAPs_equivalence}
\end{figure}

In order to ensure that the 15,000 geometries that are sampled from the WS22 database do cover all of the conformation space as spanned by the complete 120,000 geometries, a short study is performed.
UMAPs are a powerful tool to perform dimensionality reduction of data. In addition to this, they are useful tools to visualize the multidimensional feature space in ML \cite{umap}. 
To assess the chemical space coverage achieved with this form of sampling, UMAPs of the molecules can be studied. UMAPs are a dimensionality reduction method that is useful to visualize multidimensional data such as the molecular representations used in ML for QC. The resulting 2D embedding can then be used to visualize the complexity of the conformation space based on the coverage that is observed. To this end, molecular descriptors were first generated for the various molecules. The choice of descriptors for this case was the unsorted Coulomb Matrix (CM) which is calculated as:
\begin{equation}
C_{i,j}:=
    \begin{cases}
        \frac{Z_i^{2.4}}{2}~,&i=j\\
        \frac{Z_i\cdot Z_j}{\left\lVert \boldsymbol{R}_i-\boldsymbol{R}_j\right\rVert}~,&i\neq j~,
    \end{cases}
    \label{CM_eq}
\end{equation}
where, $Z_i$ is the atomic charge of the $i^{\rm th}$ atom of the molecule and $\boldsymbol{R}_i$ is its Cartesian coordinate. 
For this proxy of the conformation space that geometries cover, a 2D UMAP was generated and the resulting plots are shown in Figure \ref{fig_UMAPs_equivalence} for all 9 molecules. 
It is observed that the UMAPs for QeMFi uniformly cover the conformation space spanned by UMAPs of WS22. 
From this plot it becomes clear that the geometries sampled for the QeMFi dataset from the WS22 database uniformly cover the entire chemical space of WS22. This is true even in cases of multiple localized clusters as seen in the case of alanine or urocanic acid.
Therefore, it becomes evident that even though only 15,000 geometries are sampled from the WS22 database, these do uniformly cover the conformation space of WS22 and should therefore offer the same level of chemical complexity for ML models.

\subsection*{Single Molecule Benchmarks}
The technical validation carried out individually for SMA and o-HBDI molecules is in line with the experimental set-up of ref.~\cite{vinod23_MFML}. A total of 12,288 training samples were chosen to build the multifidelity models. 712 samples were set aside as a validation set for the o-MFML model from ref.~\cite{vinod_2024_oMFML}, and the remaining 2,000 samples were used as a test set. The accuracy of the models are gauged with mean absolute error (MAE) in the form of learning curves. Learning curves display the MAE with respect to increasing number of training samples, here, at the highest fidelity, that is, TZVP. In addition, a special kind of learning curve as seen in ref.~\cite{vinod23_MFML} are also shown. These are MAEs versus the total time to generate the training set for the MFML model. These special learning curves provide a better picture of the time-benefit of using MFML over conventional single fidelity methods. The o-MFML method additionally requires a validation set computed at the target fidelity, $F$. In these benchmarks presented herein, this is not accounted for since this work is meant as a data descriptor and not a comprehensive comparison of the MFML and o-MFML method. For the benchmarks that are reported here, the MFML and o-MFML methods perform similarly in terms of MAE. In such scenarios, it is to be concluded that MFML is better suited than o-MFML due to the lack of the cost associated with a validation set.

\begin{figure}[htb!]
    \begin{subfigure}[b]{\textwidth}
         \centering
         \includegraphics[width=0.75\textwidth]{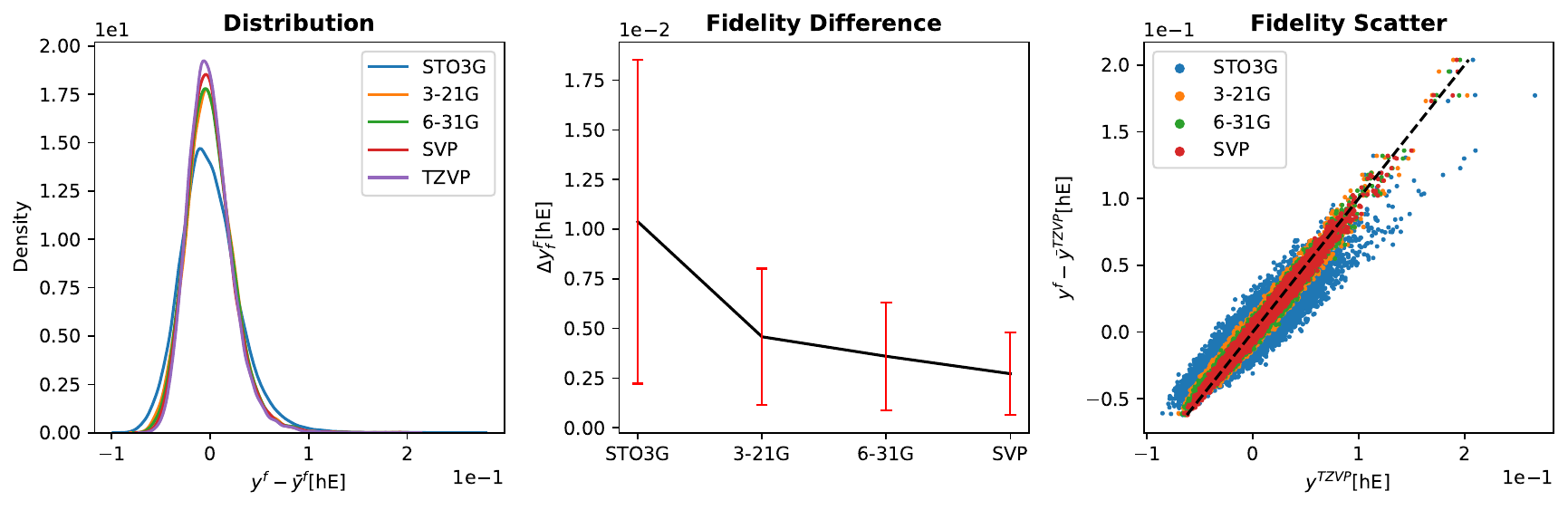}
         \caption{
         Preliminary analysis of multifidelity structure of SCF ground state energies for the SMA molecule. The three different preliminary tests for the hierarchy are performed as prescribed in ref.~\cite{vinod23_MFML}. The ground state energies show a normal distribution centered around 0 hE. The difference in the fidelity energies is monotonically decreasing for increasing fidelities indicating that the assumed hierarchy holds. The scatter plot of the energies of different fidelities with respect to the TZVP fidelity show a compact distribution for the most part. With STO3G there is a wider deviation from the identity map (dashed black line). 
         }
         \label{fig_SMA_prelim}
     \end{subfigure}
     \vfill
     \begin{subfigure}[b]{\textwidth}
         \centering
         \includegraphics[width=0.5\textwidth]{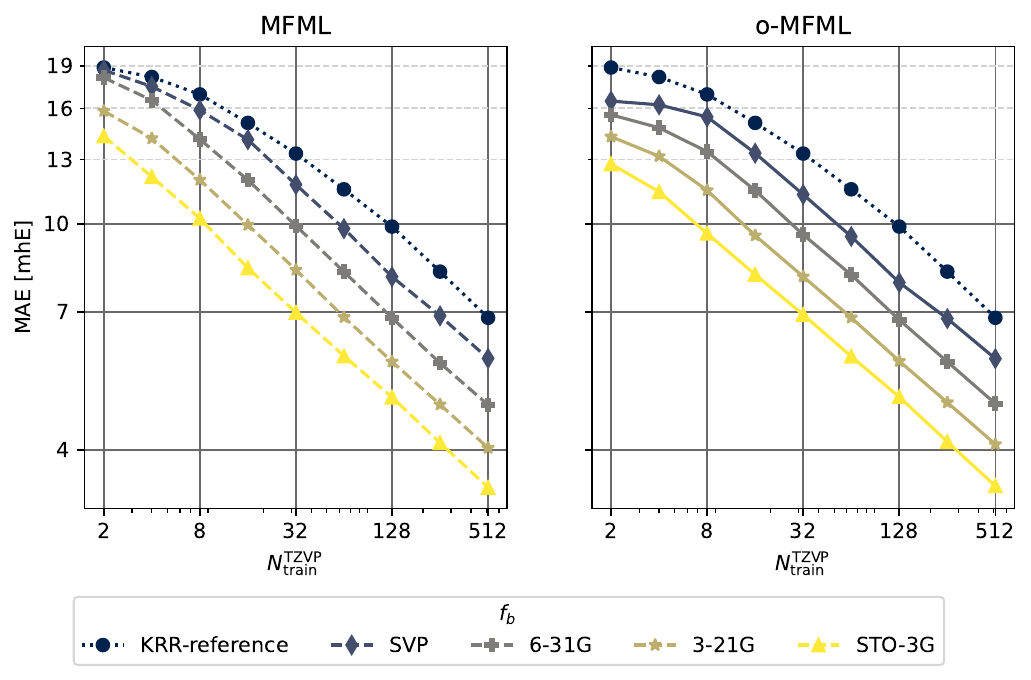}
         \caption{
         Learning curves for MFML and o-MFML for the SCF ground state energies of SMA as recorded in the QeMFi database. The reference single-fidelity KRR is also shown by training on TZVP only. The Laplacian kernel was used with a kernel width of 200.0 and regularization of $10^{-10}$. The Global SLATM \cite{Huang2020slatm} molecular descriptors were used.  
         }
         \label{fig_SMA_LC}
     \end{subfigure}
     \vfill
     \begin{subfigure}[b]{\textwidth}
         \centering
         \includegraphics[width=0.5\textwidth]{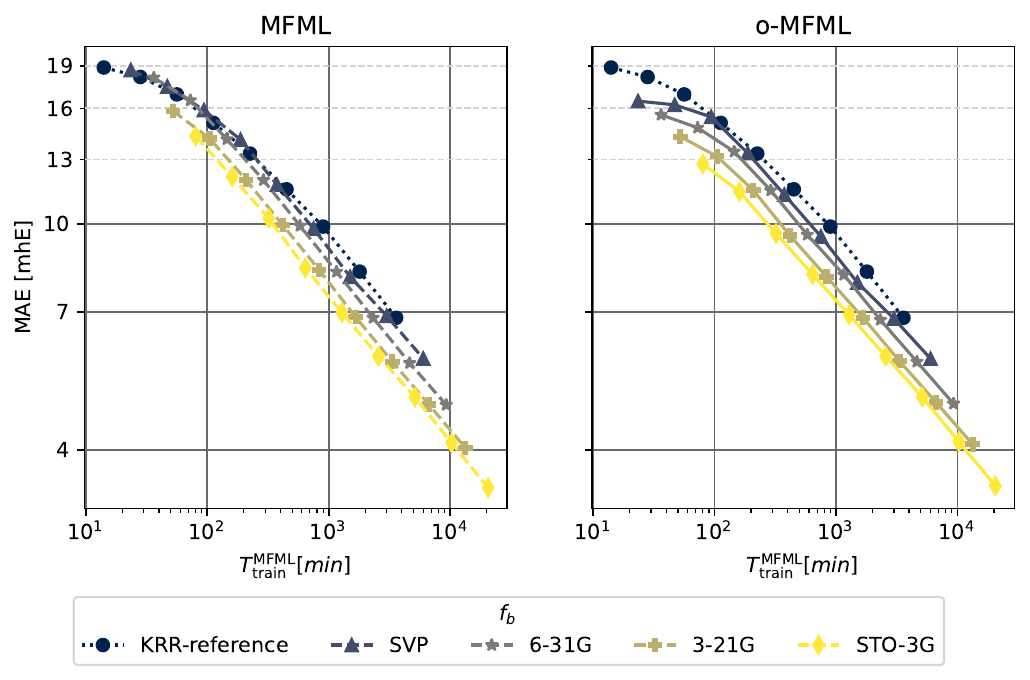}
         \caption{
         Time versus MAE plots for MFML and o-MFML models predicting the SCF ground state energies of the SMA molecule. The time to generate the training set for MFML is a comprehensive measure of the cost of a multifidelity model as prescribed in ref.~\cite{vinod23_MFML}.  
         }
         \label{fig_SMA_time_LC}
     \end{subfigure}
    \caption{Technical benchmarks of SMA from the QeMFi dataset.}
    \label{fig_SMA_benchmark}
\end{figure}

Figure \ref{fig_SMA_benchmark} reports the multifidelity benchmarking results for SCF ground state energies for SMA. The SLATM molecular descriptor \cite{Huang2020slatm} was used with a Laplacian kernel to perform kernel ridge regression. 
First, the preliminary analyses as recommended by ref.~\cite{vinod23_MFML} are shown in Figure \ref{fig_SMA_prelim}. 
The first of these is to study the distribution of the multifidelity data. The second analysis is the study of mean absolute differences between each fidelity and the target fidelity (that is, the most accurate fidelity, here, TZVP). 
Generally, it is anticipated that these differences decay monotonically for increasing fidelity.
Thirdly and finally, a scatter plot of the energies of the different fidelities with respect to the energies of the target fidelity is generated to study how these deviate with respect to the target fidelity. 
The three different preliminary analyses show a systematic ordering of the fidelities which confirm the assumed hierarchy as seen in the fidelity difference plots. 
The fidelity scatter plot also shows a systematic distribution of the energies when compared to the target fidelity of TZVP. 
In ref.~\cite{vinod23_MFML} this was identified as a good preliminary indicator of favorable results in the MFML models.
In Figure \ref{fig_SMA_LC}, the learning curves of MFML and o-MFML (with OLS optimization) are depicted for the prediction of the ground state energies. 
The multifidelity learning curves can be understood as follows:
the addition of a cheaper fidelity systematically decreases the error (here reported as mean absolute error, MAE) which is reported in units of milli-Hartree. 
With each cheaper fidelity being added, one notices that the corresponding learning curve from Figure \ref{fig_SMA_LC} has a lower offset. 
The continuing negative slope indicates that further addition of training samples could decrease the MAE of these models. There is no significant difference in MAE between the MFML and o-MFML models and they both perform similarly for the prediction of ground state energies.
As a final assessment for SMA, an MAE versus time to generate the MFML training data is also shown in Figure \ref{fig_SMA_time_LC} for the MFML and o-MFML models. The total time cost for a given MFML (or o-MFML) model is the total time taken to generate all the training samples at the different fidelities. In other words, if one picks the model $P_{\rm MFML}^{(TZVP;631G)}$ with $N_{\rm train}^{TZVP}=2$, then the total time to generate the training set for this model would be given as $T_{\rm MFML}=2\times t_{TZVP} + 4\times t_{SVP} + 8\times t_{631G}$, where $t_f$ is the computation time corresponding to the fidelity $f$. Since the QeMFi dataset comes with the average compute times for each fidelity of each molecule, this allows for a meaningful benchmarking of multifidelity models with this form of analysis. In Figure \ref{fig_SMA_time_LC} one observes that with addition of cheaper baseline fidelities, the curves show an increased offset along the time axis. Consider the very last data point of the curve corresponding to the reference KRR for o-MFML (right-hand side plot), which corresponds to $\sim 4\cdot 10^3$. If one were to draw an imaginary line parallel to the horizontal axis, it would intersect the curve corresponding to the STO3G baseline MFML model around $\sim 10^{3}$. This implies that one could use the o-MFML model with STO3G baseline to achieve similar accuracy as the reference KRR model with a reduction in time cost by a factor of $4\cdot10^3/1\cdot10^3=4$. This indeed shows the effectiveness of such forms of multifidelity models over conventional single fidelity methods.

\begin{figure}[htb!]
    \begin{subfigure}[b]{\textwidth}
         \centering
         \includegraphics[width=0.75\textwidth]{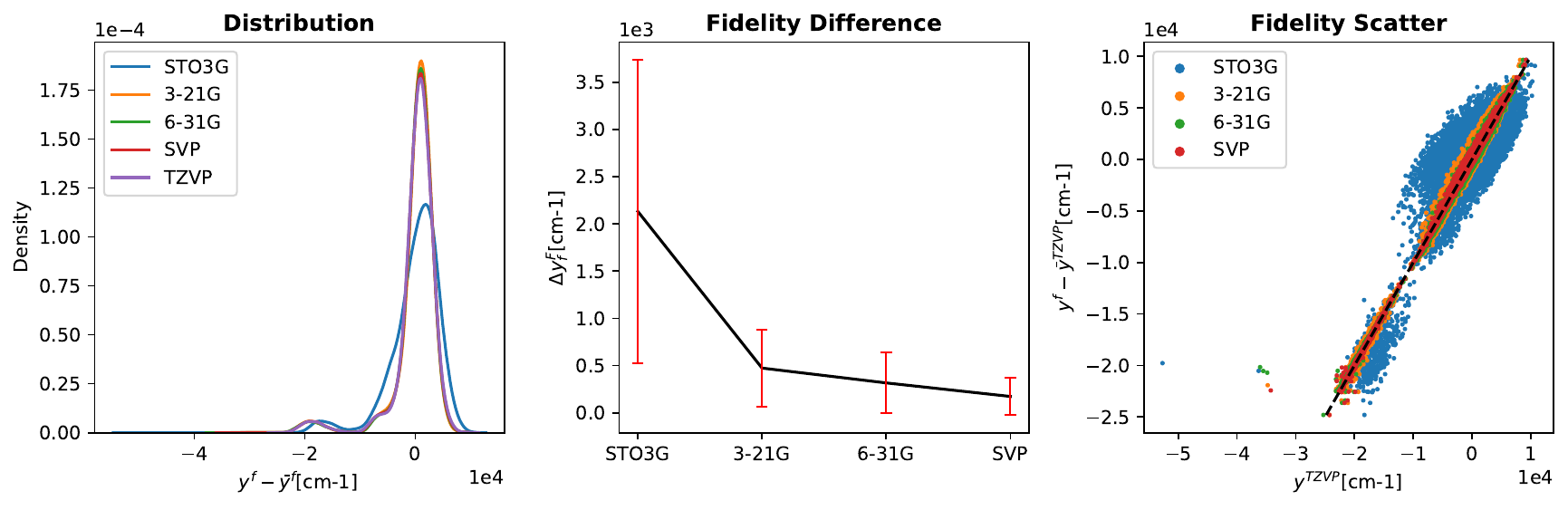}
         \caption{
         Preliminary analysis of multifidelity structure of the first vertical excitation energies for the o-HBDI molecule. The three different preliminary tests for the hierarchy are performed as prescribed in ref.~\cite{vinod23_MFML}. The different fidelities show distinct peaks around 0 $\rm cm^{-1}$ with a small bump around $18000\rm~cm^{-1}$. The difference in the fidelities shows a distinct hierarchy with reducing difference with increasing fidelity. The scatter plot of energies of different fidelities with respect to the TZVP fidelity show them well distributed except for the STO3G fidelity which shows a wider spread.
         }
         \label{fig_o-HBDI_prelim}
     \end{subfigure}
     \vfill
     \begin{subfigure}[b]{\textwidth}
         \centering
         \includegraphics[width=0.5\textwidth]{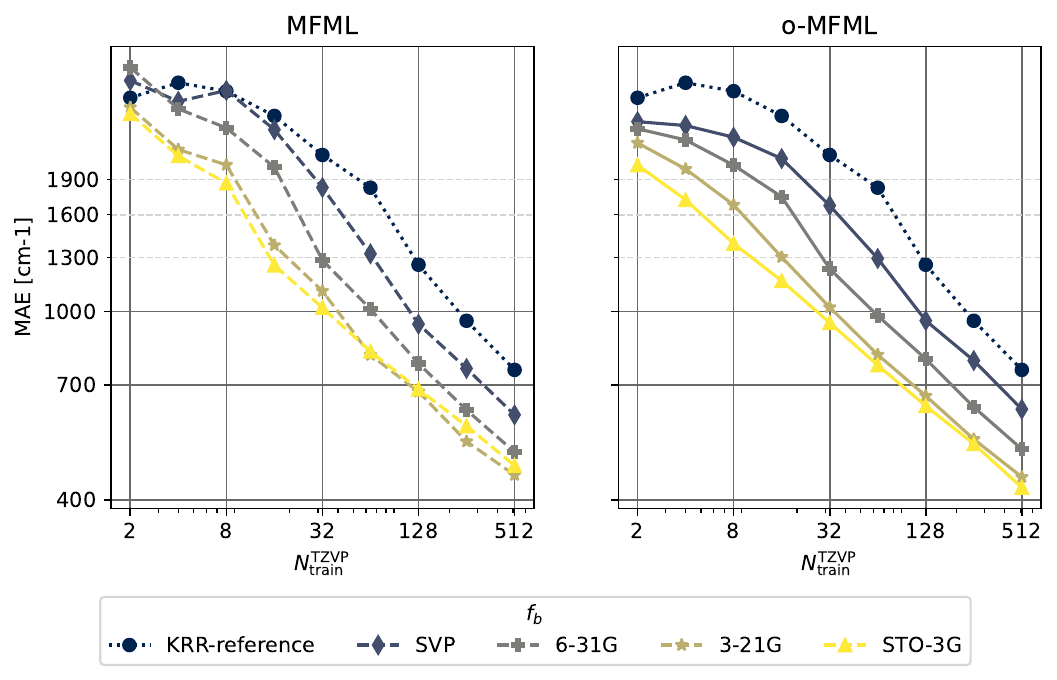}
         \caption{
         Learning curves for MFML and o-MFML for the first vertical excitation energy of o-HBDI from the QeMFi database. The reference single-fidelity KRR is also shown by training on TZVP only. The Mat\'ern kernel of first order with $L_2$-norm was used with a kernel width of 150.0 and regularization of $10^{-10}$. Unsorted CM descriptors were used for these cases.  
         }
         \label{fig_o-HBDI_LC}
     \end{subfigure}
     \vfill
     \begin{subfigure}[b]{\textwidth}
         \centering
         \includegraphics[width=0.5\textwidth]{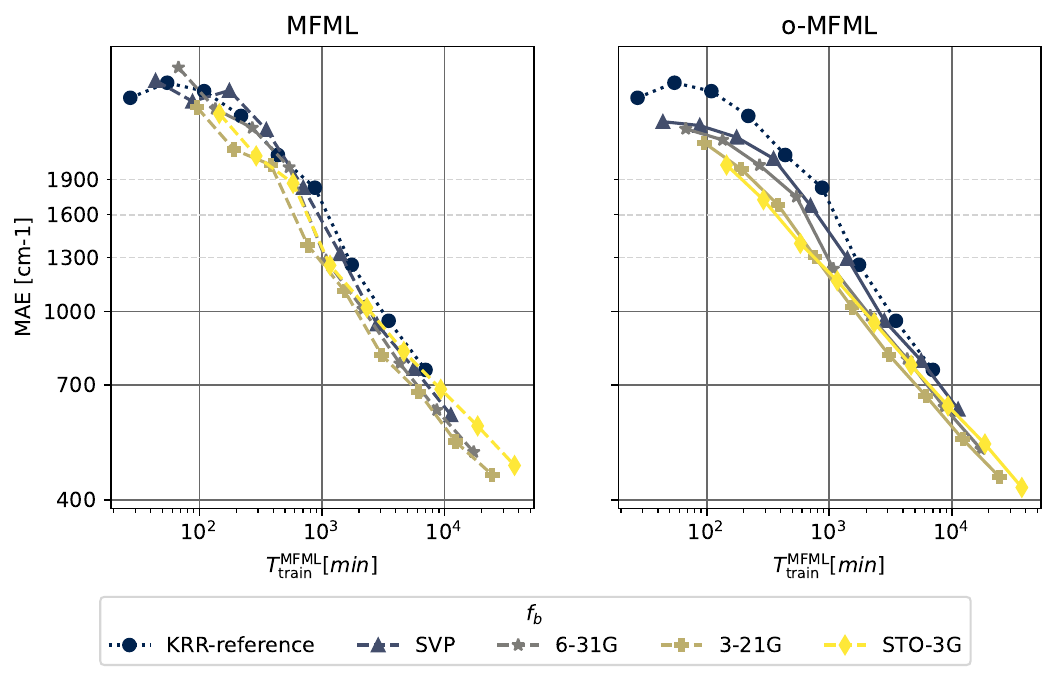}
         \caption{
         Time versus MAE plots for the prediction of first vertical excitation energy for the o-hbdi molecule. The time to generate the MFML training set is calculated as described in ref.~\cite{vinod23_MFML}.  
         }
         \label{fig_o-HBDI_time_LC}
     \end{subfigure}
    \caption{Technical benchmarks of o-HBDI from the QeMFi dataset.}
    \label{fig_o-HBDI_benchmark}
\end{figure}

A similar benchmarking procedure was carried out for the prediction of first vertical excitation energies of o-HBDI. In this case, the unsorted CM were used with the Mat\'ern Kernel. As for the case of SMA, a preliminary analysis study was performed with the resulting plots shown in Figure \ref{fig_o-HBDI_prelim}. The difference in fidelities  plot in the center indicates that the assumed hierarchy holds true for the fidelities. However, the fidelity scatter plot on the right hand side shows two distinct clusters. These correspond to the two main conformers of o-HBDI, namely the cis and trans conformers. The scatter plot also shows some cases where the STO3G fidelity covers a wider range of values and is less localized than the other fidelities. This could indicate that the use of STO3G in the MFML models would result in a lower improvement of the accuracy of the model. 

The learning curves for the prediction of the first vertical excitation energies of o-HBDI from QeMFi are shown in Figure \ref{fig_o-HBDI_LC}. The MAE are reported in $\rm cm^{-1}$ with the axes identically scaled for both MFML and o-MFML. With the addition of cheaper fidelities, the learning curves show a constant reduction in the offset of the MAE as seen in the near parallel learning curves of the different baselines fidelities. As anticipated from the preliminary analysis, the addition of the STO3G fidelity does not provide significant improvement especially for larger training samples. However, this is rectified, as expected, by the o-MFML method which was indeed shown to fix this very issue in ref.~\cite{vinod_2024_oMFML}. Indeed, in the MAE versus time to generate training data plots seen in Figure \ref{fig_o-HBDI_time_LC} for o-HBDI, one observes that the STO3G baseline model fails to provide a reasonable improvement in the time benefit unlike observed for the case of SMA in Figure \ref{fig_SMA_time_LC}. However, for models built with with the other baseline fidelities, a time improvement is still visible. These results are a strong indicator towards the possibility of further research in the field of multifidelity methods for QC.

\subsection*{Cumulative use of the dataset}
The QeMFi dataset contains multifidelity QC properties of 9 molecules for 15,000 geometries. This totals to $9\times 15,000=135,000$ point calculations of the QC properties. 
This is therefore the largest collection of multifidelity dataset which can be used in various benchmarking processes.
To demonstrate this form of cumulative use of the dataset, multifidelity models from ref.~\cite{vinod_2024_oMFML} were tested against this in predicting the ground state energies of the molecules.
From each molecule of the QeMFi dataset, 1,500 geometries were randomly chosen and compiled into a total of $9\times 1,500=13,500$ data points.  
From the 13,500 samples, a random set of 11,000 samples were used as the multifidelity training data. 
Of the remaining, 500 samples were used as a validation set and 2,000 as the holdout test set. 
With this setup learning curves were generated for the different multifidelity models in the same fashion as prescribed in ref.~\cite{vinod_2024_oMFML}.

\begin{figure}[htb!]
    \centering
    \includegraphics[width=\textwidth]{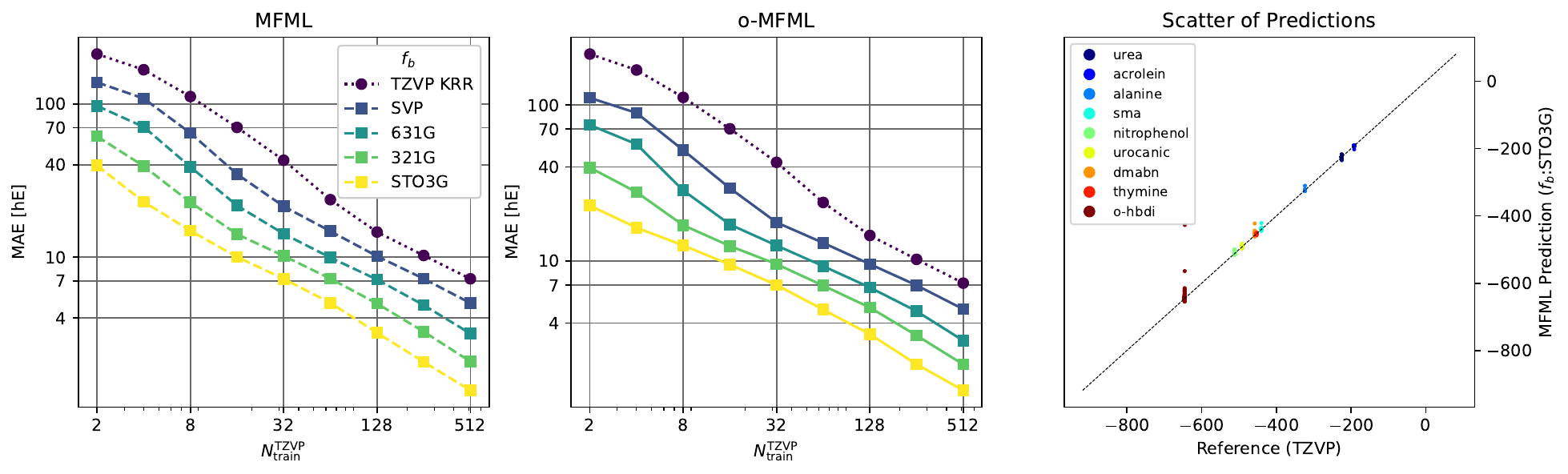}
    \caption{Learning curves for MFML and o-MFML for SCF ground state energies based on the cumulative use of the QeMFi dataset. 1,500 samples were randomly chosen from each molecule to perform this example test. The single fidelity KRR is also shown for comparison. The scatter plot of MFML predicted versus reference ground state energies is also shown. The MFML and o-MFML models perform well on the cumulative dataset showing errors in the range of a few hE for the ground state energies.}
    \label{fig_extended_dataset_LC}
\end{figure}

The results of this test are shown in Figure \ref{fig_extended_dataset_LC} for MFML and o-MFML models. The learning curves show a decreasing slope for both cases for the different baseline fidelities. The addition of each cheaper fidelity results in a lower offset of MAE. The constant slope on the log-log axis indicates that addition of training samples can further decrease the MAE. 
On the right-hand side plot the scatter of reference TZVP versus MFML predicted SCF ground state energies are delineated. Across the energy ranges the MFML model predicts the SCF ground state energies accurately as can be inferred from the scatter of the values being close to the identity mapping line. 

\begin{figure}[htb!]
    \centering
    \includegraphics[width=0.8\textwidth]{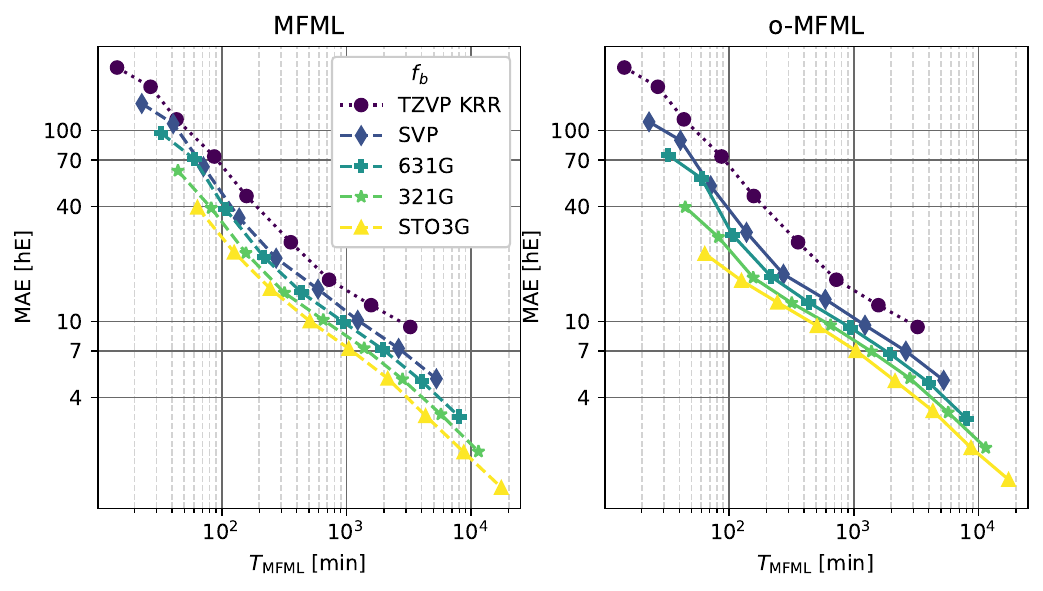}
    \caption{MAE versus time taken to generate a training set for MFML and o-MFML for the cumulative use of the QeMFi dataset. 
    Here one observes the time-benefit of using MFML over single fidelity KRR for each cheaper baseline fidelity.}
    \label{fig_extended_dataset_timeMAE}
\end{figure}

For completeness, the MAE versus time to generate the multifidelity training data are reported in Figure \ref{fig_extended_dataset_timeMAE} for MFML and o-MFML for this cumulative use of the QeMFi dataset.
In this specific case for the cumulative use of the QeMFi dataset, the time to generate the multifidelity training data is calculated based on the molecular geometry that is included in the model. 
It is observed in Figure \ref{fig_extended_dataset_timeMAE}, that the addition of each baseline fidelity results in a distinct reduction in the time cost of generating training data. 
The resulting curves indicate a time benefit factor of about 6 for the STO3G baseline fidelity. 

\section*{Usage Notes}
In addition to the multifidelity dataset, various tools to assess and benchmark multifidelity methods are also provided. 
These include scripts to perform preliminary analysis of the data based on the property of choice as recommended in ref.~\cite{vinod23_MFML}, and the scripts to produce learning curves. Further, scripts to generate unsorted Coulomb Matrices, and the global SLATM descriptors are provided which are built upon the qmlcode package \cite{Christensenqmlcode}. 
The scripts are easy to use and well documented allowing for a streamlined benchmarking process with an example shown in Listing~\ref{lst_prelimanalysis} for the preliminary analysis. Listing~\ref{lst_MFML_LCs} shows an example to generate the multifidelity learning curves.

\section*{Code availability}
All scripts needed to assess this dataset are hosted at \href{https://github.com/SM4DA/QeMFi}{https://github.com/SM4DA/QeMFi}. This includes scripts to run ORCA calculations, extract properties from the output log files, generating CM and SLATM molecular descriptors, and generating learning curves.

\bibliography{sample}

\section*{Acknowledgements}
The authors acknowledge support by the DFG through projects ZA 1175/3-1 as well as through the DFG Priority Program SPP 2363 on “Utilization and Development of Machine Learning for Molecular Applications – Molecular Machine Learning” through the project ZA 1175\_4-1. 
The computations of the quantum chemical properties were carried out on the \href{https://pleiades.uni-wuppertal.de}{PLEIADES cluster} at the University of Wuppertal, which was supported by the Deutsche Forschungsgemeinschaft (DFG, grant No. INST 218/78-1 FUGG) and the Bundesministerium für Bildung und Forschung (BMBF). The authors would also like to thank the `Interdisciplinary Center for Machine Learning and Data Analytics (IZMD)' at the University of Wuppertal.

\section*{Author contributions statement}
V.V.: Conceptualization, data collection, methodology, validation, writing - original draft preparation, review \& editing. 
\newline
P.Z.: Funding acquisition, supervision, validation, writing - review \& editing

\section*{Competing interests}
The authors declare that there is no conflict of interest or any competing interests. 

\section*{Figures \& Tables}

\begin{table}[htb!]
    \centering
    \begin{tabular}{|l|c|c|}
    \hline
         \textbf{PROPERTY}& \textbf{DIMENSIONS/UNITS} & \textbf{npz ID}\\
         \hline
         \hline
         Atomic Numbers$^\dag$ & \texttt{(n\_atoms,)} & \texttt{`Z'}\\
         Cartesian Coordinates$^\dag$ & \texttt{(n\_atoms,)}/\AA & \texttt{`R'}\\
         Ground State Energies (SCF) & \texttt{(15000, 5)}/hE & \texttt{`SCF'}\\
         Vertical Excitation Energies & \texttt{(15000,5,10)}/$\rm cm^{-1}$ & \texttt{`EV'}\\
         Transition Dipole Moments & \texttt{(15000,5,10,3)}/a.u. & \texttt{`TrD'}\\
         Oscillator Strength & \texttt{(15000,5,10)} & \texttt{`fosc'}\\
         Molecular Dipole Moment (electronic) & \texttt{(15000,5,3)}/a.u. & \texttt{`DPe'}\\
         Molecular Dipole Moment (nuclear) & \texttt{(15000,5,3)}/a.u. & \texttt{`DPn'}\\
         Rotational Constants & \texttt{(15000,5,3)}/cm-1 & \texttt{`RCo'}\\
         Dipole Moment Along Rotational Axis & \texttt{(15000,5,3)}/a.u. & \texttt{`DPRo'}\\
         QC Calculation Times  & \texttt{(5,)}/seconds & \texttt{`t'}\\
         \hline
    \end{tabular}
    \caption{List of properties available in the QeMFi dataset. The corresponding dimension(s) and units of the properties are also given with the npz file key. $^\dag$From the WS22 database \cite{ws22_data_paper, ws22_dataset_zenodo}}
    \label{tab_properties_list}
\end{table}

\begin{lstlisting}[float, language=Python, caption={Python example to extract the SVP fidelity values of second vertical excitation state of alanine from QeMFi.},frame=single, numbers=right,label={lst_codeaccess}]
import numpy as np

#load the dataset for alanine
data = np.load(`QeMFi_alanine.npz')
#query for the vertical excitation energies
EV = data[`EV']
#Select the second vertical state for SVP (4th fidelity)
EV_SVP = EV[:,3,1]

#load QC compute times
QC_time = data[`t']
\end{lstlisting}

\begin{lstlisting}[float,language=bash, caption={Python example to perform preliminary analysis of SCF ground state multifidelity data for the SMA molecule.}, frame=single, numbers=right, label={lst_prelimanalysis}]
# this code is running in an activated conda environment
(conda) $ python PrelimAnalysis.py -m=`sma' \
-d=`../dataset/' \
-p=`SCF' \
-u=`hE' \
--centeroffset \
--saveplot
\end{lstlisting}

\begin{lstlisting}[float, language=bash, caption={Python example to generate the global SLATM representations, multifidelity learning curves, and the corresponding plot of SCF ground state energies for the SMA molecule.}, numbers=right,frame=single,label={lst_MFML_LCs}]
# this code is running in an activated conda environment

#this creates the SLATM representations
(conda) $ python GenerateSLATM.py -m=`sma' \ 
-d=`../dataset/' 

#this runs the code to generate learning curves of MFML
(conda) $ python LearningCurves.py -m=`sma' \
-d=`../dataset/' \
-p=`SCF' \
-n=10 \
-w=200.0 \
-rep=`CM' \
-k=`laplacian' \
-r=1e-10 \
--seed=42 \
--centeroffset 

#this creates the various MAE files in npy format
#the following command will plot the learning curves as a PDF file
(conda) $ python LC_plots.py -m=`sma' \
-p=`SCF' \
-u=`hE' \
-rep=`CM' \
--centeroffset \
--saveplot
\end{lstlisting}

\end{document}